\documentclass{PoS}
\usepackage{slashed}
\title{The milliQan experiment: search for milli-charged particles at the LHC}

\ShortTitle{Milliqan update}

\author{\speaker{Haitham Zaraket}\thanks{Work partially supported by the Lebanese University Grant. On behalf of the MilliQan Collaboration.}\\
        Lebanese University, Faculty of Sciences, Multidisciplinary Physics Lab \\
        E-mail: \email{hzaraket@ul.edu.lb}}

\abstract{Charge quantization has always been enigmatic. Theoretically, Millicharged particles can 
be an answer. The use of existing detectors, without affecting their initial 
mandate, is a very promising low cost new physics detector for millicharged particles. 
The Milliqan collaboration 
has installed a 1/100th version of the full detector at LHC point 5 in CMS. Data 
collected in 2018 is now under investigation/analysis. Some results are presented 
in this paper.}

\FullConference{
European Physical Society Conference on High Energy Physics - EPS-HEP2019 -\\
			10-17 July, 2019\\
			Ghent, Belgium}

\begin{document}

\section{Why milli-charged?}
Charge quantization is a remarkable feature of the standard model (SM). Stable particles with charge < ~0.3e (quarks already known to have charges of 1/3e and 
2/3e) are not part of the predicted and detected flora of particles in the SM. 
Millicharge particles, if they exist, are invisible to current LHC  
detection system due to their low charge. The low charge implies low energy deposits (single photons) in exiting particle detectors and are highly affected by particle background.

The search for millicharged particle is related to complementary searches at 
LHC (not covered by ATLAS/CMS searches) for hidden/dark sector. For example 
massless dark photons are related in several models to the existence of such millicharged particles and can have distinctive signature at the LHC. 
From the astrophysical side,  
recent excitement has shown that milli-charged
dark matter at the 1$\%$ mass-density level is a leading explanation \cite{MunozL} for the EDGES 21-cm result \cite{BowmaRMMM}. Constraints from astrophysical observation on the small couplings to the Standard Model can be 
used to guide our search for millicharged particles (for instance \cite{ChangEM}). 

Millicharged particles can arise in several models. Among the simplest scenarios 
one can imagine, a simple extension to the Standard Model can be made by just 
adding a new U(1) gauge symmetry. The coupling to the standard model can then be made by 
adding a kinetic mixing term among the gauge invariant field tensors $F_{\mu\nu}$ (for ${\rm U}_{EM}(1)$) and $B_{\mu\nu}$ (for the additional ${\rm U}^\prime(1)$)
\begin{equation}
{\cal L}={\cal L}_{SM}-\frac{\kappa}{4}B_{\mu\nu}F^{\mu\nu}- \frac{1}{4}B_{\mu\nu}B^{\mu\nu} \; .
\end{equation}
A new fermion fermion is then added with coupling to   ${\rm U}^\prime(1)$: 
\begin{equation}
{\cal L}={\cal L}_{SM}-\frac{\kappa}{4}B_{\mu\nu}F^{\mu\nu}- \frac{1}{4}B_{\mu\nu}B^{\mu\nu}+\bar{\psi}\left(i\slashed{\partial}-e^\prime\slashed{B}-m\right)\psi\; . 
\end{equation}
The above theory is equivalent to the theory where we can make the change of field 
$ B_\mu\rightarrow B_\mu +\kappa A_\mu$. Hence eliminating the coupling between the 
photon field $A_\mu$ and the dark photon ($B_\mu$), but generating a new charge to the new fermion $\kappa e^\prime$ (the millicharged particle mCP):
\begin{equation}
{\cal L}={\cal L}_{SM}- \frac{1}{4}B_{\mu\nu}B^{\mu\nu}+\bar{\psi}\left(i\slashed{\partial}-\kappa e^\prime\slashed{A}-e^\prime\slashed{B}-m\right)\psi \; .
\end{equation}   
Models with additional symmetry groups can also be used to generate millicharged particles. 
Additional massive bosons ($Z^\prime$) can be considered beside dark photons. 

The main millicharged particle production mechanisms are QCD inspired processes. In pp collisions, besides Drell-Yan, we can list processes with $\eta$, $\eta^\prime$, $\rho$ and $J/\psi$ decay. The detection  can be 
either through the decay of the mCP or by scattering, energy loss.  

\section{Detector proposal}
There is a long history of direct and indirect searches for mCPs. For mCPs with mass 
below that of the electron one finds strong bounds from astrophysical observations and cosmological models. 
Laboratory tests can be highlighted ranging from indirect 
decay of ortho-positronium to that of accelerator beam experiment (SLAC for instance \cite{SLAC}). 
  
The challenge in detecting a low charge particle comes from the fact that we have 
a lower ionization energy. A detector with large depth is needed for the particle 
to traverse. Segmentation is needed to be sure that the passing particle is an 
ionizing particle. The idea of the MilliQan detector came in 2014 \cite{Milliqan1} to add detector sensitive 
to milli-charged particles produced in LHC collisions with:  
\begin{itemize}
\item a charge down to $10^{-3}$e, hence an energy loss of $10^{-6}$ lower than that 
of an e-charged particle is expected. Hence the 
need for long, sensitive active length to see a single photo-electron. 
\item The proposal was a 1 m x 1 m x 3 m scintillator plus photomultiplier 
Tube (PMT) array, pointing back to interaction point, in well shielded area near CMS 
or ATLAS. 
\item With triple coincidence, dominant random background should be controlled.  
\end{itemize}
In 2016 a Letter Of Intent (LOI) was published \cite{LOI} identifying the location in CMS point 5 and a first Full detector simulation was made.

\section{Status of Demonstrator}
\begin{figure}
\centering{\includegraphics[width=.6\textwidth]{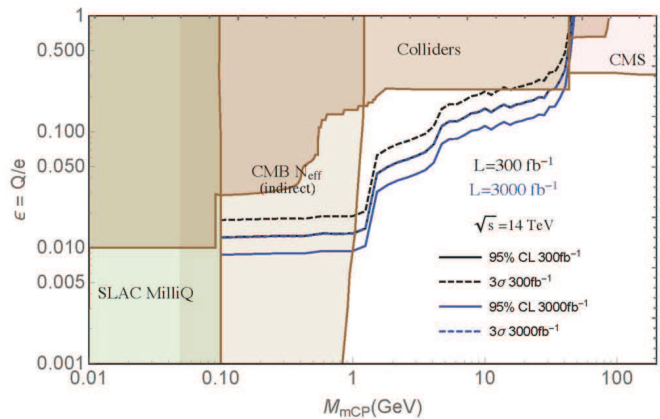}}
\caption{The simulated sensitivity with 15 ns coincidence interval with three layers of the detector.}
\label{Sensitivity}
\end{figure}

In the fall of 2017 a $1\%$ scale "demonstrator" of the proposed detector was installed in CMS drainage gallery at 33~m from the CMS IP. The main 
mission of the demonstrator was to study
the feasibility of the experiment, focusing on understanding various background sources such as
radioactivity of materials, PMT dark current (main challenge for single PE), cosmic rays, and beam induced backgrounds. Being around 70~m underground a natural, but not, 
complete shield from cosmic muons is expected. The expected sensitivity is shown in 
Figure~\ref{Sensitivity}. An updated sensitivity curve taking into account a realistic model 
of the detector and its installation position besides full simulation of background 
is under investigation by the collaboration.  

\begin{figure}
\centering{\includegraphics[width=.6\textwidth]{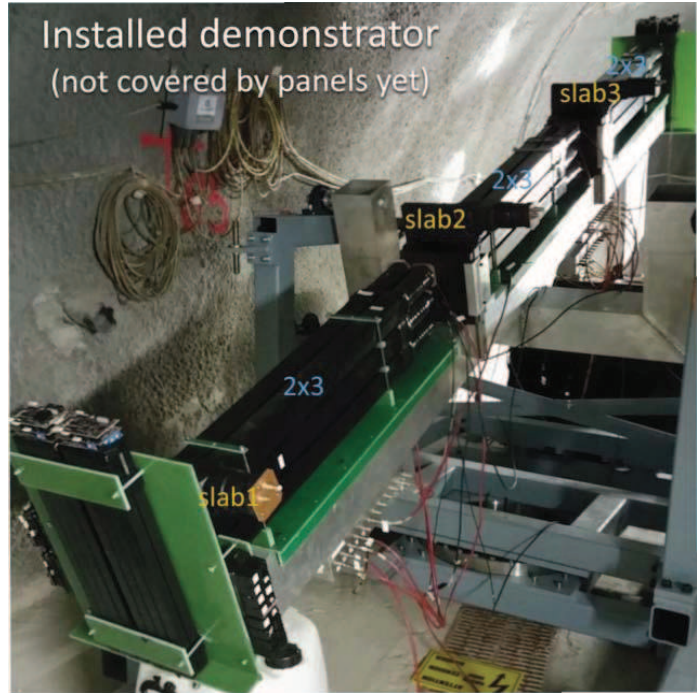}}
\caption{The initial installation of the $1\%$ of the MilliQan detector.}
\label{Demo1}
\end{figure}

\begin{figure}
\centering{\includegraphics[width=.6\textwidth]{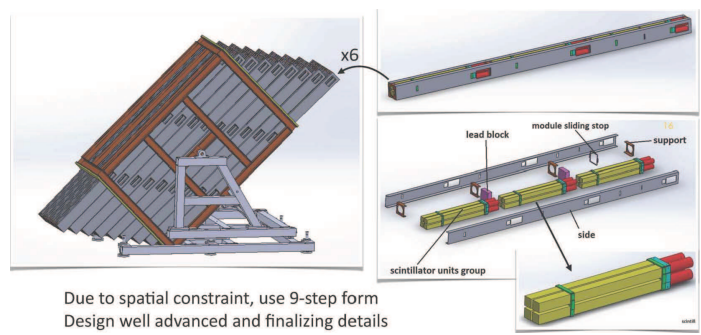}}
\caption{The proposed Modular mechanical design of the full MilliQan detector.}
\label{Demo2}
\end{figure}

The initial design respected the LOI proposal by implementing three layers of 2x3 scintillator plus PMT. Several PMT species were used.  Events are triggered if there is a signal above the trigger threshold 
in three trigger groups within a window of $100$~ns. Each trigger group contains two channels and the trigger
group for each channel is given by floor(channel/2). The channel mapping ensures adjacent bars are in the same trigger group. The trigger thresholds varied during data taking to satisfy rate requirements. Scintillator slabs and lead bricks were added to analyze background and Tag through-going particles or to shield radiation. There are also several hodoscope packs composed of small arrays of $0.75 \times 18$ inch rectangular pieces of plastic scintillator readout via SiPMs attached at one 
 end. These provide finer grained position information that allows crude tracking through the device (see Figure~\ref{Demo1}). 
As a sign of readiness for full detector installation the used support structure was designed to 
hold the full detector. Besides, a full mechanical design was made with a modular 
strategy (see Figure~\ref{Demo2}) to simplify the upgrade and maintenance of the detector, keeping our promise of not affecting any activity of CMS.   

\section{Highlights from Demonstrator Data}
In this section we give some highlights of the analysis of the data taken during the 
last year. Full analysis of the result will be published later by the collaboration. 

\begin{figure}
\centering{\includegraphics[width=.6\textwidth]{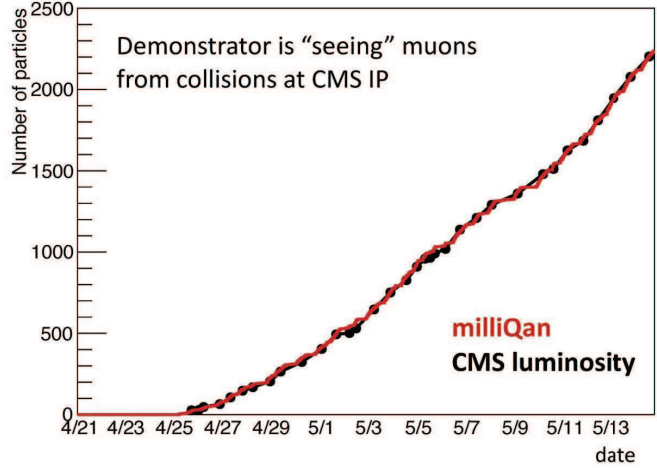}}
\caption{The luminosity measured by CMS compared to the rate of through-going 
particles through the MilliQan detector. }
\label{lumino}
\end{figure}

In 2018 run a very successful 37~fb$^{-1}$ was done with around  more than 2000 h trigger live-time. A very valuable experience in operating the detector was performed. In December 2018 two additional channels were added to the design as a fourth layer. The run and collected data was shown to be a powerful resource to study and optimize the  performance of final detector. 
A first test for the correct alignment of the detector was made by comparing the rate of through-going particles during a fill and comparing it with the luminosity time constant (14 h in this case) as shown in Figure~\ref{lumino}. 

Cosmic Muons and Muons from CMS were detected and the data taken was validated with the simulated 
GEANT efficiency. An {\it in-situ} ${\rm N}_{PE}$ per charge calibration was done. It should be mentioned that the calibration of the ${\rm N}_{PE}$ per unit charge incident on the 
detector is crucial for determining the lowest charge to which milliQan can be sensitive. 
To achieve this, we first compute the number of PEs for cosmic muons 
incident on the demonstrator. The value for ${\rm N}_{PE}$ is extracted by dividing the pulse area of cosmic 
muons by the pulse area of a single PE obtained from delayed scintillation PEs. The 
method of using delayed scintillation PEs to measure the SPE response was validated 
 using  an LED bench measurement. Extrapolating the ${\rm N}_{PE}$  to fractional charges by scaling by $Q^2$ an estimation of ${\rm N}_{PE}=1$ was obtained for 
 $Q\sim 3\times 10^{-3}e$.  
Triggering is another important mission of the prototype to be tested and validated.  

\begin{figure}
\centering{\includegraphics[width=.6\textwidth]{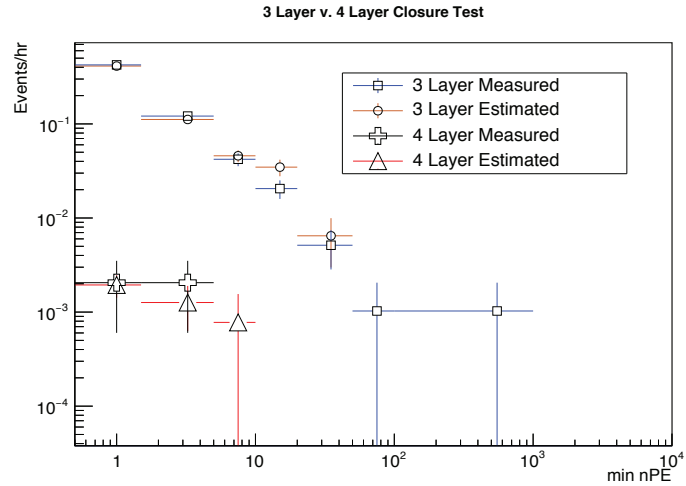}}
\caption{The ABCD closure test with three and four layers comparing data and simulation. }
\label{ABCD}
\end{figure}

The total background as a function of the minimal ${\rm N}_{PE}$ per charge in the event can be evaluated 
using data taking periods in which there are no collisions and scaled to the 
total length of the data taking with collisions.
To study further correlated backgrounds we used the ABCD method [A: pointing path with coincidence less than a threshold (15 ns or 30 ns), B: non pointing path with coincidence less than the threshold,  C: non pointing path with coincidence greater than the threshold, D: pointing path with coincidence greater than the threshold). As can be shown in the closure test 
shown in Figure~\ref{ABCD} perfect matching is observed between data and simulation. 
In December 2018 a fourth layer was added to the installed demonstrator. The data and the simulation has 
shown a  reduction of  the background by two orders of magnitude compared to three layers design as can be seen in Figure~\ref{ABCD}. 

\section{Summary}
The search for millicharged particles using existing accelerators facilities represents a very promising path. The feasibility prototype for MilliQan (the 
demonstrator) has operate successfully the last year with very positive signs 
for full detector installation. With the data collected triggering, time and charge 
calibration were possible.  A clear understanding of background was performed.

\end{document}